\documentclass[iop,twocolumn,showpacs,showemail]{revtex4}
\usepackage{bbm}
\usepackage{mathrsfs}
\usepackage{epsfig,psfrag}
\usepackage{amsmath,amsfonts,amssymb}
\usepackage[usenames]{color}
\usepackage[dvips, bookmarks, colorlinks=true, plainpages = false, citecolor = blue, linkcolor = blue, urlcolor = blue, filecolor = blue]{hyperref}
\def\be{\begin{equation}}
\def\ee{\end{equation}}
\def\bea{\begin{eqnarray}}
\def\eea{\end{eqnarray}}

\begin{document}

\preprint{draft}

\title{Fractal Structure of Hastings$-$Levitov Patterns Restricted in a Sector Geometry}

\author{F. Mohammadi$^1$}\author{A. A. Saberi$^2$}\email{a$_$saberi@ipm.ir} \author{S. Rouhani$^1$ }

\address {$^1$ Department of Physics, Sharif University of Technology, P.O. Box 11155-9161,
Tehran, Iran. \\$^2$ School of Physics, Institute for Research in
Fundamental Sciences (IPM), P.O.Box 19395-5531, Tehran, Iran. }
\date{\today}

\pacs{61.43.Hv, 82.40.Ck, 68.43.Jk, 64.60.al}

\begin{abstract}
A generalized form of the Hastings and Levitov (HL) algorithm for
simulation of diffusion-limited aggregation (DLA) restricted in a
sector geometry is studied. It is found that this generalization
with uniform measure produces "wedge-like" fractal patterns in the
physical space, whose fractal dimension and anisotropy exponent
depend significantly on the opening angle $\beta$ of the sector. The
morphological properties and the overall shape of the patterns are
analyzed by computing the angular two-point density correlation
function of the patterns. We also find that the fractal dimension of
the patterns with sinusoidal distributed measure depend weakly on
$\beta$ with almost the same dimension as the radial DLA cluster.
The anisotropy exponent and the visual appearance of the patterns in
this case are shown to be compatible with those of the
advection-diffusion-limited aggregation (ADLA) clusters.
\end{abstract}

\maketitle
\section{introduction}

Diffusion-limited aggregation (DLA) was originally introduced by
Witten and Sander \cite{Ref1} in 1981 to model the aggregates of
metal particles formed by adhesive contact in low concentration
limit. This model has been then shown to describe many pattern
forming processes including dielectric breakdown \cite{Ref2},
electrochemical deposition \cite{Ref3, Ref4}, viscous fingering and
Laplacian growth \cite{Ref5}. One of the standard approaches to
simulate a Laplacian field is by random walkers, which are launched
from the periphery of the system and diffuse toward the growing
cluster and freeze on it. This procedure is equivalent to solving
Laplace equation outside the aggregated cluster with an appropriate
boundary conditions. The walker sticks to a point on the surface of
the aggregate with a probability proportional to the harmonic
measure there \cite{Ref5-0}.\\Another powerful method for studying
such growth processes in two-dimensions, is the iterated stochastic
conformal mapping \cite{Ref5-1, Ref5-2, Ref5-3}, which is known as
Hastings and Levitov (HL) method.

Since DLA enhances the instability of growth, the resulting cluster
is highly ramified and branched. The procedure of the proliferation
of the fingers in the fractal structure of DLA is one of its
complexity sources which is one of the purposes of the present paper
to deal with.

The HL method is based on the fact that there exists a conformal map
that maps the exterior of the unit circle in the mathematical
$w$-plane to exterior of the cluster of $n$ particles in the
physical $z$-plane. The complicated harmonic measure in the physical
space then changes to a simple measure: a uniform distribution in
the mathematical space. The HL algorithm can be also used for more
generalized growth models for which the probability measure is not
uniform in the mathematical plane. In a part of this paper, we also
study the statistical and morphological properties of the fractal
patterns produced by a novel generalization of the HL algorithm with
both uniform and non-uniform probability measures in the $w$-plane.
We restrict the algorithm to attribute the measure in the $w$-plane
to a sector of opening angle $\beta$. We find that this restriction
which breaks the radial symmetry in the growth process, yields
fractal patterns whose dimensions depend on whether uniform or
non-uniform distribution is used for the measure. For the uniform
distribution, the patterns looks like those of DLA in a wedge but
with fractal dimension depending on the opening angle $\beta$.
However for the non-uniform distribution of the measure, we find
that the statistical properties of the patterns is compatible with
advection-diffusion-limited aggregation (ADLA) \cite{Ref5-4} and
their fractal dimensions take values close to that of radial DLA
i.e., $d_f=1.71$ with a weak dependence on the opening angle
$\beta$.

Bazant \emph{et al.} \cite{Ref5-4} proposed a method based on
time-dependent conformal maps to model a class of non-Laplacian
growth processes such as ADLA in a background potential flow. They
showed that in spite of dramatic increases in anisotropy, the
fractal dimension of the patterns is not affected by advection and
takes the same value as for radial DLA clusters.

The fractal dimension of the patterns is already shown to be
affected by the geometrical factors. Stepanov and Levitov
\cite{Ref6} obtained dimensions as low as $d_f=1.5$ for simulations
of anisotropic growth (a different anisotropy that we consider in
this paper) using the noise-controlled HL algorithm. A similar
result reported in \cite{Ref7} for DLA with anisotropic
perturbations. Davidovitch \emph{et al.}, \cite{Ref8} introduced
growth models that are produced by deterministic itineraries of
iterated conformal maps with different dimensions. In channel
geometry, the fractal dimensions $d_f=1.67$ \cite{Ref9} and
$d_f=1.71$ \cite{Ref10} are reported for periodic and reflecting
boundary conditions,
respectively.\\
The comparison of the ensemble average of DLA cluster density with
the noise-free Saffman-Taylor viscous fingers \cite{Ref5} have been
the subject of various studies in both channel and wedge geometries
\cite{Ref11, Ref12, Ref13, Ref14}.

There has been shown in \cite{Ref15} that there exists a critical
angle $\eta\approx60^\circ-70^\circ$ in viscous fingers and DLA
growing in a wedge, indicative of a typical angular spread of a
major finger. In this paper, for the patterns produced by restricted
HL algorithm with uniform measure, we find that their visual
appearance like a DLA cluster in a wedge and we compute the average
angle between two major fingers. To estimate the angle, we measure
the angular density-density correlation function. We also obtain the
relation between the minimum opening angle of a wedge which contains
the cluster in the $z$-plane, and the opening angle $\beta$
considered in the $w$-plane.

\section{Restricted HL algorithm in a sector geometry\label{RHL}}

Hastings and Levitov (HL) employed the conformal-mapping tool to
describe the complicated boundary of a growing DLA cluster. Their
approach \cite{Ref5-1} was based on iteratively applying the
function $\phi_{\lambda , \theta}(w)$ which maps a unit circle to a
circle with a bump of linear size $\sqrt{\lambda}$ at the point $w =
e^{i\theta}$,
\begin{equation*}
    \phi_{\lambda , 0}(w) =     w^{1-a}    \left\{\frac{1+\lambda}{2w}(1+w) \right.
\end{equation*}
\vspace{-20 pt}
\begin{equation}\label{Eq4}
    \left.   \left[1+w+w\left(1+\frac{1}{w^2}-\frac{2}{w}\frac{1-\lambda}{1+\lambda}    \right)^\frac{1}{2} \right]
  -1\right\}^a,
\end{equation}
\begin{equation}\label{Eq5}
    \phi_{\lambda , \theta}(w)=e^{i\theta}\phi_{\lambda
    ,0}\left(e^{-i\theta}w\right).
\end{equation}
The parameter $0\leq a\leq1$ determines the shape of the bump, for
higher $a$ the bump becomes elongated in the normal direction to the
boundary $\partial \mathcal{C}$ (e.g., it is a line segment for
$a=1$). In this paper we set $a=\frac{1}{2}$ for which the bump has
a semi-circle shape.

A cluster $\mathcal{C}_n$ consisting of $n$ bumps can be obtained by
using the following map on a unit circle
\begin{equation}\label{Eq6}
    \Phi_{n}(w) = \phi_{\lambda_{1},\theta_{1}} \circ
    \phi_{\lambda_{2},\theta_{2}} \circ \cdots \circ
    \phi_{\lambda_{n},\theta_{n}} (w),
\end{equation}
which corresponds to the following recursive relation for a cluster
$\mathcal{C}_{n+1}$,
\begin{equation}\label{Eq7}
    \Phi_{n+1}(w) = \Phi_{n}(\phi_{\lambda_{n+1},\theta_{n+1}}(w)).
\end{equation}
In order to have bumps of fixed-size on the boundary of the cluster,
since the linear dimension at point $w$ is proportional to
$|\Phi^{'}_{n}(w)|^{-1}$, one obtains
\begin{equation}\label{Eq9}
    \lambda_{n+1} =
    \frac{\lambda_{0}}{|\Phi^{'}_{n}(e^{i\theta_{n+1}})|^2}.
\end{equation}
To produce an isotropic radial DLA, the harmonic measure of the
$n$-th growth probability $p(z,n)$ has to be conformally mapped onto
a constant measure $p(\theta_n)=\frac{1}{2\pi}$ on a unit circle
i.e., $0\leq \theta_n \leq 2\pi$ (more details of the simulation
algorithm which have been also considered in this paper are given in
our previous work \cite{Ref16}).

In this paper we use a generalized form of the HL algorithm in which
$\theta_n$s are restricted to be selected from a fixed interval
$0\leq \theta_n \leq \beta$ on a unit circle, with both uniform and
non-uniform measures. This measure
\begin{equation}\label{Eq}
                \left\{ \begin{array}{cl}
                p(\theta_n) &  0\leq \theta_n \leq \beta\\
                0 & \textrm{otherwise},
              \end{array}\right.
\end{equation}
induces a complicated harmonic measure on the boundary of the
patterns in the $z$-plane.

\section{Scaling Properties of the Produced Fractal Patterns\label{Scaling}}

The function $\Phi_{n} (w)$ in Eq. ({¬\ref{Eq6}}), has the following
Laurent series expansion
\begin{equation}\label{Eq9m0}
    \Phi_{n} (w)= F_1^{(n)}w+F_0^{(n)}+F_{-1}^{(n)}w^{-1}+\cdot\cdot\cdot ,
\end{equation}
where $F_1^{(n)}$ is called the \emph{conformal radius} and
$F_0^{(n)}$ the \emph{center of charge} of the cluster
$\mathcal{C}_n$. It is possible to analytically show that
\cite{Ref5-3} these coefficients contain descriptive information
about the morphology of the clusters. In particular, for the
coefficient of the linear term in Eq. (\ref{Eq9m0}), it can be shown
that \cite{Ref5-3}
\begin{equation}\label{Eq9mm0}
   F_1^{(n)} = \prod_{k=1}^{n}(1+\lambda_k )^a,
\end{equation}
which scales with the cluster size $n$ as
\begin{equation}\label{d_f}
   F_1^{(n)} \sim n^{1/d_f}\sqrt{\lambda_0},
\end{equation}
where $d_f$ denotes the fractal dimension of the cluster.

Scaling behavior of the next Laurent coefficient $F_0^{(n)}$ can
also explain about the isotropicity of the clusters. For an
isotropic DLA cluster, this coefficient scales according to the
following relation
\begin{equation}\label{d_0}
   |F_0^{(n)}| \sim n^{1/{d_0}},
\end{equation}
with $2/d_0=0.7$ \cite{Ref5-3}. This also holds for the anisotropic
growth phenomena but with a different exponent $d_0$, e.g., $d_0\sim
1.71$ is obtained for ADLA \cite{Ref5-4}.\\Since our produced
patterns are anisotropic, in order to have a quantitative measure
and compare them with that of the other studied models, we check the
scaling relation (\ref{d_0}), and compute the exponent $d_0$ as a
function of the opening angle $\beta$ (for convenience, we address
$d_0$ as \emph{anisotropy exponent}, in this paper). We compute
$F_0^{(n)}$ by using the recursion equation given in \cite{Ref5-3}
for the Laurent coefficients.

\section{Restricted HL patterns with Uniform Measure\label{RHL-UM}}

\begin{figure}[!htb]
\includegraphics[width=0.47\textwidth]{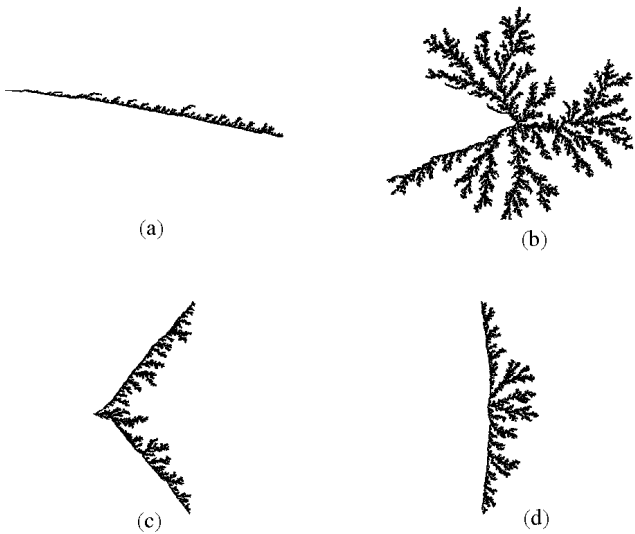}
\caption{Some typical clusters of size $n= 10^5$, generated by using
the restricted HL algorithm with uniform measure and (a) $\beta =
36^\circ$, (b) $\beta =324^\circ$, (c) $\beta = 144^\circ$ and (d)
$\beta =216^\circ$.} \label{fig1-}
\end{figure}

\begin{figure}[!htb]
\includegraphics[width=0.48\textwidth]{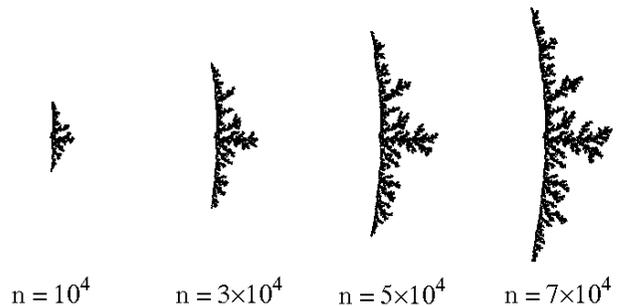}
\caption{Snapshots of a growing cluster of $\beta =216^\circ$ at
different number of iterative conformal maps $n$.} \label{fig1-a}
\end{figure}

In this section, we investigate the effect of the restriction made
on the angular distribution of the measure in the mathematical
plane, on the statistical properties of the patterns produced by
using the algorithm described in section \ref{RHL}, with
$p(\theta_n)=\frac{1}{\beta}$ and $0\leq\theta_n \leq \beta$.\\
We grew $400$ clusters of size $n=10^4$, and $30$ clusters of size
$n=10^5$ for different opening angles in the range $4.5^\circ\leq
\beta \leq 360^\circ$. Having looked at the sample patterns shown in
Fig. {¬\ref{fig1-}}, the appearance of a wedge-like shape in the all
patterns is evident. Each of the two interval limits for the
distribution of $\theta_n$s, i.e., $\theta_n = 0$ and $\beta$, can
represent itself as a boundary of the wedge (which due to the
conformal maps, are not necessarily straight lines) in the physical
plane. Despite that the harmonic measure in $w$-plane is uniform,
the growth probability looks much larger around the boundaries in
the $z$-plane. For small angles of $\beta \lesssim 36^\circ$ there
always appears only one finger with sidebranches on a boundary. For
larger $\beta$ there also exists at least one finger which has
selected a boundary to grow on. In the range angle that two main
branches coexist, because of the attraction of the boundaries, the
average angle between branches increases by increasing $\beta$.\\
The snapshots of an example of growing cluster with $\beta
=216^\circ$ in Fig. \ref{fig1-a} graphically show the procedure of
the branch formation in the growing fractal pattern.

\subsection{Scaling exponents}

\begin{figure}[t]
\includegraphics[width=0.47\textwidth]{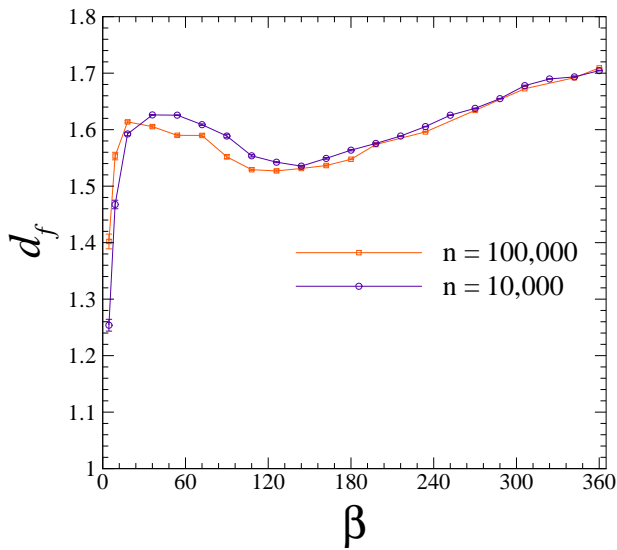}
\caption{The fractal dimension $d_f$ of the patterns produced by
using the restricted HL algorithm with uniform measure as a function
of the opening angle $\beta$ in $w$-plane. The averages were taken
over $400$ clusters of size $n=10^4$ (circles) and $30$ clusters of
size $n=10^5$ (squares).} \label{fig2-}
\end{figure}

To have a quantitative understanding of the behavior of our
clusters, let us now measure some of their statistical quantities.
The first quantity we would like to measure is the fractal dimension
which can be obtained using the scaling relation in Eq. (\ref{d_f}).
Two sets of results, one averaged over $400$ clusters of size $10^4$
and the other over $30$ clusters of size $10^5$ at each opening
angle $\beta$, are reported in Fig. \ref{fig2-}. The fractal
dimension shows a significant dependence on the opening angle
$\beta$. For narrow angles the dimension takes values very close to
unity which is dominated by the boundary effects. By increasing
$\beta$, it reaches a relative maximum at around $\beta \simeq 18
^\circ$ which is almost the half-maximum angle interval in which
there exists only one main branch. There also exists a local minimum
of the dimension around $\beta \simeq 144 ^\circ$ which seems to be
a typical angle up to which only two main branches coexist. For
greater angles, the dimension increases until for $\beta=360^\circ$
where an isotropic radial DLA with $d_f=1.71$ is expected. For DLA
clusters in a wedge produced by the original definition \cite{Ref1},
the fractal dimension is not affected by the wedge geometry of
opening angle $\alpha$ \cite{Ref15}, and it depends weakly, if at
all, on $\alpha$.

\begin{figure}[h]
\includegraphics[width=0.47\textwidth]{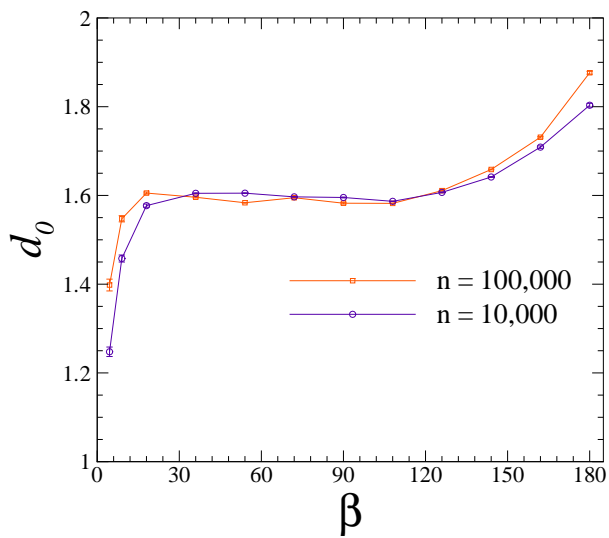}
\caption{The exponent $d_0$ as a function of $\beta$ in the scaling
region, according to Eq. (\ref{d_0}).} \label{fig3-}
\end{figure}

The other quantity which we measure for our clusters is the exponent
describing the scaling behavior of the position of the \emph{center
of charge} $|F_0^{(n)}|$ with the cluster size $n$, defined in Eq.
(\ref{d_0}). We find such a scaling behavior only in the interval
$9^\circ\leq \beta \leq 180^\circ$. There has been also observed
that the corresponding exponent $d_0$ depends on $\beta$ in this
range. The results for two mentioned sets of averages are shown in
Fig \ref{fig3-}.

\subsection{Overall shape and the morphology of the clusters}

In this subsection we discuss about the overall shape and the
morphology of the clusters. As mentioned before, for sectors of
angle $\beta\lesssim 36 ^\circ$ there exists only one main branch,
and for greater angles up to $\beta\simeq 144 ^\circ$ there are two
main coexistent branches, and for angles $\beta\gtrsim 144 ^\circ$
more than two main branches appear. To quantify this observation and
obtain the average angle between two main branches of the patterns,
we follow the same reasoning as in \cite{Ref15}, for our clusters.
Kessler \emph{et al.}, \cite{Ref15} studied the building block of
DLA clusters in a wedge by computing the angular two-point density
correlation function in the constitutive sectors of a cluster.

\begin{figure}[h]
\includegraphics[width=0.25\textwidth]{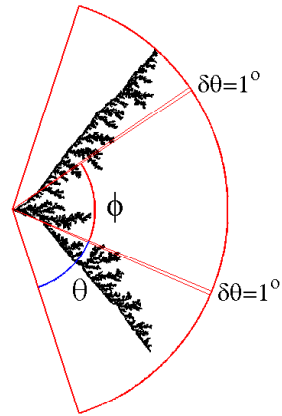}
\caption{Schematic description of the procedure expressed in the
text to compute the angular correlation function according to Eq.
(\ref{correlation}).} \label{fig4-a}
\end{figure}

\begin{figure}[t]
\includegraphics[width=0.44\textwidth]{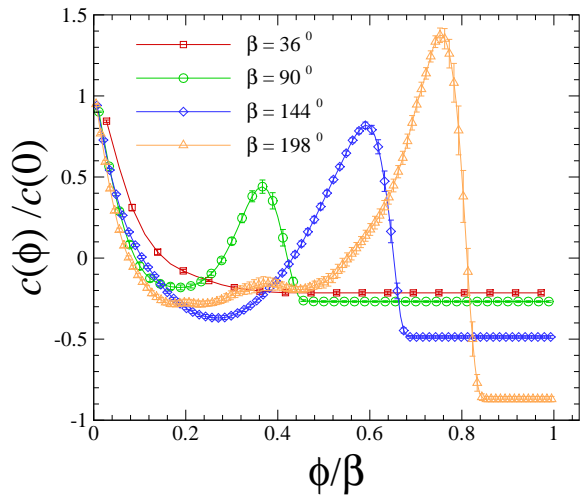}
\caption{Angular correlation function $c(\phi) / c(0)$ for clusters
produced by using the restricted HL algorithm with uniform measure
for different opening angle $\beta$ as a function of $\phi/\beta$.}
\label{fig4-}
\end{figure}

Consider two sectors separated by angle $\phi$. The density-density
correlation function is then read off from
\begin{equation}\label{correlation}
   c(\phi) = [\langle\rho(\theta+\phi) \rho(\theta)\rangle - \langle\rho\rangle^2]|_\theta
   \, ,
\end{equation}
where $\rho(\theta)$ is the density of particles in the cluster in a
$\delta\theta=1^{\circ}$ sector around $\theta$ (see Fig.
\ref{fig4-a}), and $\langle\cdot\cdot\rangle$ denotes for averaging
over the sectors in $z$-plane. The computed correlation functions
for different opening angle $\beta$ averaged over $30$ realizations
are shown in Fig. \ref{fig4-}. As can be seen in the figure, for
clusters consisting of one main branch there is an anticorrelation
between the origin and other angles. Appearance of the second peak
in the function with positive correlation is indicative of the
second coexistent main branch in the cluster.

Location of the second peak shows approximately the angle between
two main coexisting branches $\eta$ in the physical plane which is
reported in the second column of Table \ref{table1}.

For a given opening angle $\beta$ in $w$-plane, the produced
patterns have a wedge-like shape in the $z$-plane. To have an
estimate relation between two angles $\beta$ and the average opening
angle of the wedge-like patterns in the $z$-plane $\alpha$, we
report some of these corresponding angles in the third column of
Table \ref{table1}. The averages are taken over $30$ clusters of
size $n=10^5$ for each reported $\alpha$.

\begin{table}[!h]
  \centering
  \begin{tabular}{|c||c|c|}
  \hline
  $\beta$ & $\eta$ & $\alpha$\\
  \hline\hline
  $36^{\circ}$ & there is only one main branch & $15^{\circ}$\\
  \hline
  $90^{\circ}$& $31^{\circ} - 34^{\circ}$ &  $44^{\circ}$\\
  \hline
  $144^{\circ}$ & $81^{\circ} - 88^{\circ}$ &  $97^{\circ}$\\
  \hline
  $198^{\circ}$& more than two main branches exist &  $164^{\circ}$\\
  \hline
\end{tabular}
\caption{The average angle between two main branches $\eta$, and the
average opening angle of the wedge-like patterns in the $z$-plane
$\alpha$ for different opening angle $\beta$ in the $w$-plane.
}\label{table1}
\end{table}

\section{Restricted HL patterns with Sinusoidal Measure\label{RHL-Sin}}

\begin{figure}[t]
\includegraphics[width=0.47\textwidth]{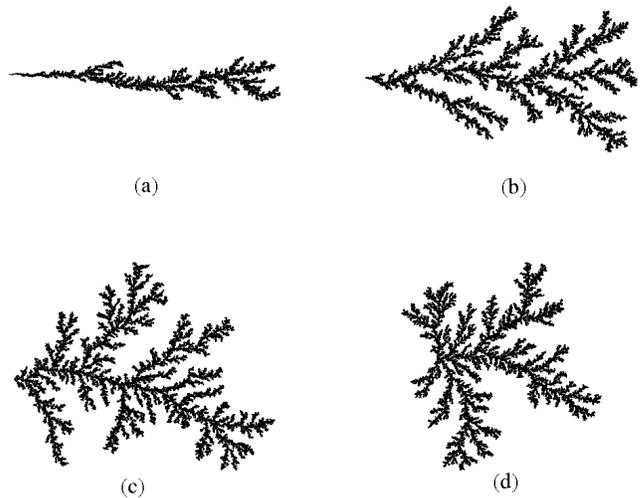}
\caption{Some typical clusters of size $n= 10^5$, generated by using
the restricted HL algorithm with sinusoidal measure i.e.,
$\sin(\pi\theta_n/\beta)$ distribution and (a) $\beta = 90^\circ$,
(b) $\beta = 180^\circ$, (c) $\beta = 270^\circ$ and (d) $\beta =
360^\circ$.} \label{fig5-}
\end{figure}

As discussed in section \ref{RHL-UM}, the boundary effects are
evident in the visual appearance and statistical properties of the
patterns. So this motivated us to force the measure to be
distributed away from the boundaries i.e., $\theta_n=0$ and $\beta$
in the $w$-plane. By inspiration from \cite{Ref5-4}, we produced
patterns according to the algorithm described in section \ref{RHL},
with $p(\theta_n)\sim\sin (\pi \theta_n / \beta)$ for $0\leq\theta_n
\leq \beta$. We generated $400$ clusters of size $n=10^4$, and some
of size $n=10^5$ for different opening angles in the range
$36^\circ\leq \beta \leq 360^\circ$. Some of the clusters are shown
in Fig. {¬\ref{fig5-}}. There exists a significant difference
between the overall shape of the patterns in figures \ref{fig1-} and
\ref{fig5-}. Due to the non-uniform measure, the visual appearance
of the patterns in Fig. \ref{fig5-} is not affected by the boundary.

\begin{figure}[h]
\includegraphics[width=0.47\textwidth]{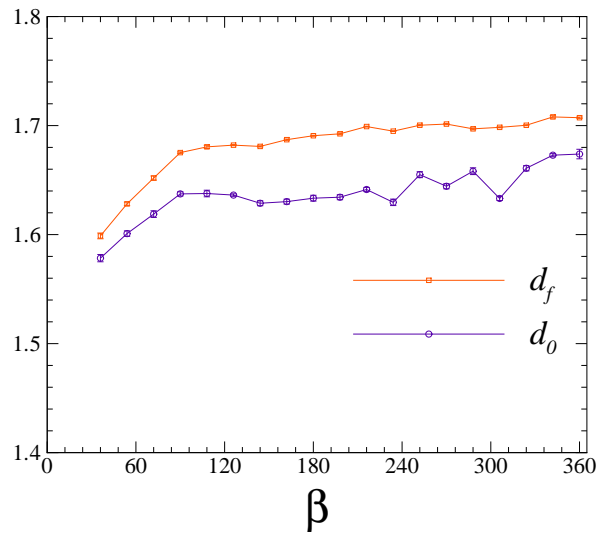}
\caption{The fractal dimension $d_f$ and the scaling exponent $d_0$
as a function of $\beta$, computed for the patterns generated by
using the restricted HL algorithm with sinusoidal measure.}
\label{fig6-}
\end{figure}

This difference can be also observed in the fractal dimension of the
patterns. As plotted in Fig. {¬\ref{fig6-}}, except for small
opening angles, the fractal dimension depends weakly on $\beta$ and
takes values close to $d_f \simeq 1.7$.

Due to the sinusoidal distribution of the measure, even for
$\beta=360^\circ$ the clusters are anisotropic. The growth process
is dominated by advection toward a certain direction with a spatial
extent which depends on the opening angle $\beta$. The morphology of
these patterns are very similar to that of ADLA clusters.\\ADLA is
one of the simplest examples of transport-limited aggregation (TLA)
in which the released random walkers are being drifted in the
direction of a background potential flow. The difference between
simulations of TLA and DLA by HL algorithm is in the sequences of
the angles $\theta_n$. In TLA the angles are chosen from a
time-dependent (non-harmonic) measure $p(\theta, t_n)$.

We have also examined the scaling relation Eq. (\ref{d_0}) for our
clusters with sinusoidal measure and obtained the dependence of the
exponent $d_0$ on the opening angle $\beta$. As shown in Fig.
\ref{fig6-}, $d_0$ is weakly dependent on $\beta$ with values of a
little less than $d_f$ which is roughly in agreement with the same
scaling behavior of ADLA which is reported in \cite{Ref5-4} for a
special case $\beta=360^\circ$.

\section{Conclusions}

The fractal structure and statistical properties of the patterns
generated by a generalized HL method restricted in a sector geometry
were studied. It is found that the restriction with uniform measure
leads to the production of fractal patterns whose fractal dimension
and anisotropy exponent depend significantly to the opening angle
$\beta$ of the sector in the mathematical $w$-plane. The overall
shape of these patterns is governed by a wedge-like DLA appearance
which has been shown to be the characteristic feature of the uniform
measure. The boundary effects lead to a nontrivial dependence of the
average angle between two major coexistence branches on the angle
$\beta$. The proliferation of the main branches was studied by
computing the angular density-density correlation function giving a
quantitative understanding of the morphology of the patterns as a
function of $\beta$.\\ It is also found that the sinusoidal
distribution of the measure on the sectors gives fractal patterns
with almost the same fractal dimension for all values $\beta$. The
anisotropy exponent and the visual appearance of these patterns are
shown to behave very similar to those of ADLA clusters.

\textbf{Acknowledgement.} A.A.S acknowledges financial support from
INSF grant.


\pagebreak


\begin{thebibliography}{00}


\expandafter\ifx\csname
natexlab\endcsname\relax\def\natexlab#1{#1}\fi
\expandafter\ifx\csname bibnamefont\endcsname\relax
  \def\bibnamefont#1{#1}\fi
\expandafter\ifx\csname bibfnamefont\endcsname\relax
  \def\bibfnamefont#1{#1}\fi
\expandafter\ifx\csname citenamefont\endcsname\relax
  \def\citenamefont#1{#1}\fi
\expandafter\ifx\csname url\endcsname\relax
  \def\url#1{\texttt{#1}}\fi
\expandafter\ifx\csname urlprefix\endcsname\relax\def\urlprefix{URL
}\fi \providecommand{\bibinfo}[2]{#2}
\providecommand{\eprint}[2][]{\url{#2}}



\bibitem[{\citenamefont{Witten, T.A. et~al.}(1981)}]{Ref1}
  \bibinfo{author}{\bibfnamefont{T.A.}~\bibnamefont{Witten}} and
  \bibinfo{author}{\bibfnamefont{L.M.}~\bibnamefont{Sander}},
  \bibinfo{Jurnal}{{Phys. Rev. Lett.}}
  \textbf{\bibinfo{volume}{47}},
  \bibinfo{pages}{1400}
 (\bibinfo{year}{1981}).


\bibitem[{\citenamefont{Niemeyer et~al.}(1984)}]{Ref2}
  \bibinfo{author}{\bibfnamefont{L.}~\bibnamefont{Niemeyer}},
  \bibinfo{author}{\bibfnamefont{L.}~\bibnamefont{Pietronero}},
  \bibinfo{author}{\bibfnamefont{H.J.}~\bibnamefont{Wiesmann}},
  \bibinfo{Jurnal}{{Phys. Rev. Lett.}}
  \textbf{\bibinfo{volume}{52}},
  \bibinfo{pages}{1033}
 (\bibinfo{year}{1984}).


\bibitem[{\citenamefont{Brady et~al.}(1984)}]{Ref3}
  \bibinfo{author}{\bibfnamefont{R.M.}~\bibnamefont{Brady}} and
  \bibinfo{author}{\bibfnamefont{R.C.}~\bibnamefont{Ball}},
  \bibinfo{Jurnal}{{Nature (London)}}
  \textbf{\bibinfo{volume}{309}},
  \bibinfo{pages}{225}
 (\bibinfo{year}{1984}).


\bibitem[{\citenamefont{Matsushita et~al.}(1984)}]{Ref4}
  \bibinfo{author}{\bibfnamefont{M.}~\bibnamefont{Matsushita}},
  \bibinfo{author}{\bibfnamefont{M.}~\bibnamefont{Sano}},
  \bibinfo{author}{\bibfnamefont{Y.}~\bibnamefont{Hayakawa}},
  \bibinfo{author}{\bibfnamefont{H.}~\bibnamefont{Honjo}},
  \bibinfo{author}{\bibfnamefont{Y.}~\bibnamefont{Sawada}},
  \bibinfo{Jurnal}{{Phys. Rev. Lett.}}
  \textbf{\bibinfo{volume}{53}},
  \bibinfo{pages}{286}
 (\bibinfo{year}{1984}).


\bibitem[{\citenamefont{Paterson et~al.}(1984)}]{Ref5}
  \bibinfo{author}{\bibfnamefont{L.}~\bibnamefont{Paterson}}
  \bibinfo{Jurnal}{{Phys. Rev. Lett.}}
  \textbf{\bibinfo{volume}{52}},
  \bibinfo{pages}{1621}
 (\bibinfo{year}{1984}).


\bibitem[{\citenamefont{}()}]{Ref5-0}
  \bibinfo{author}{\bibfnamefont{A.A.}~\bibnamefont{Saberi}},
  \bibinfo{Jurnal}{{J. Phys.: Condens. Matter}}
  \textbf{\bibinfo{volume}{21}},
  \bibinfo{pages}{465106}
 (\bibinfo{year}{2009}).


 \bibitem[{\citenamefont{}()}]{Ref5-1}
  \bibinfo{author}{\bibfnamefont{M.B.}~\bibnamefont{Hastings}} and
  \bibinfo{author}{\bibfnamefont{L.S.}~\bibnamefont{Levitov}},
  \bibinfo{Jurnal}{{Physica D}}
  \textbf{\bibinfo{volume}{116}},
  \bibinfo{pages}{244}
 (\bibinfo{year}{1998}).

\bibitem[{\citenamefont{}()}]{Ref5-2}
  \bibinfo{author}{\bibfnamefont{M.B.}~\bibnamefont{Hastings}}
  \bibinfo{Jurnal}{{Phys. Rev. E}}
  \textbf{\bibinfo{volume}{55}},
  \bibinfo{pages}{135}
 (\bibinfo{year}{1997}).


\bibitem[{\citenamefont{}()}]{Ref5-3}
  \bibinfo{author}{\bibfnamefont{B.}~\bibnamefont{Davidovitch}},
  \bibinfo{author}{\bibfnamefont{H.G.E.}~\bibnamefont{Hentchel}},
  \bibinfo{author}{\bibfnamefont{Z.}~\bibnamefont{Olami}},
  \bibinfo{author}{\bibfnamefont{I.}~\bibnamefont{Procaccia}},
\bibinfo{author}{\bibfnamefont{L.M.}~\bibnamefont{Sander}},
\bibinfo{author}{\bibfnamefont{E.}~\bibnamefont{Somfai}},
  \bibinfo{Jurnal}{{Phys. Rev. E}}
  \textbf{\bibinfo{volume}{59}},
  \bibinfo{pages}{1368}
 (\bibinfo{year}{1999}).



 \bibitem[{\citenamefont{}()}]{Ref5-4}
  \bibinfo{author}{\bibfnamefont{M.Z.}~\bibnamefont{Bazant}},
  \bibinfo{author}{\bibfnamefont{J.}~\bibnamefont{Choi}},
  \bibinfo{author}{\bibfnamefont{B.}~\bibnamefont{Davidovitch}},
  \bibinfo{Jurnal}{{Phys. Rev. Lett.}}
  \textbf{\bibinfo{volume}{91}},
  \bibinfo{pages}{045503}
 (\bibinfo{year}{2003}).


\bibitem[{\citenamefont{}()}]{Ref6}
  \bibinfo{author}{\bibfnamefont{M.G.}~\bibnamefont{Stepanov}} and
  \bibinfo{author}{\bibfnamefont{L.S.}~\bibnamefont{Levitov}},
  \bibinfo{Jurnal}{{Phys. Rev. E}}
  \textbf{\bibinfo{volume}{63}},
  \bibinfo{pages}{061102}
 (\bibinfo{year}{2001}).


\bibitem[{\citenamefont{}()}]{Ref7}
  \bibinfo{author}{\bibfnamefont{M.N.}~\bibnamefont{Popescu}},
  \bibinfo{author}{\bibfnamefont{H.G.E.}~\bibnamefont{Hentschel}},
  \bibinfo{author}{\bibfnamefont{F.}~\bibnamefont{Family}},
  \bibinfo{Jurnal}{{Phys. Rev. E}}
  \textbf{\bibinfo{volume}{69}},
  \bibinfo{pages}{061403}
 (\bibinfo{year}{2004}).


\bibitem[{\citenamefont{}()}]{Ref8}
  \bibinfo{author}{\bibfnamefont{B.}~\bibnamefont{Davidovitch}},
  \bibinfo{author}{\bibfnamefont{M.J.}~\bibnamefont{Feigenbaum}},
  \bibinfo{author}{\bibfnamefont{H.G.E.}~\bibnamefont{Hentschel}},
  \bibinfo{author}{\bibfnamefont{I.}~\bibnamefont{Procaccia}},
  \bibinfo{Jurnal}{{Phys. Rev. E}}
  \textbf{\bibinfo{volume}{62}},
  \bibinfo{pages}{1706}
 (\bibinfo{year}{2000}).


\bibitem[{\citenamefont{}()}]{Ref9}
  \bibinfo{author}{\bibfnamefont{B.B.}~\bibnamefont{Mandelbrot}},
  \bibinfo{author}{\bibfnamefont{A.}~\bibnamefont{Vespignani}} and
  \bibinfo{author}{\bibfnamefont{H.}~\bibnamefont{Kaufman}},
  \bibinfo{Jurnal}{{Europhys. Lett.}}
  \textbf{\bibinfo{volume}{32}},
  \bibinfo{pages}{199}
 (\bibinfo{year}{1995}).



\bibitem[{\citenamefont{}()}]{Ref10}
  \bibinfo{author}{\bibfnamefont{E.}~\bibnamefont{Somfai}},
  \bibinfo{author}{\bibfnamefont{R.C.}~\bibnamefont{Ball}},
  \bibinfo{author}{\bibfnamefont{J.P.}~\bibnamefont{DeVita}},
  \bibinfo{author}{\bibfnamefont{L.M.}~\bibnamefont{Sander}},
  \bibinfo{Jurnal}{{Phys. Rev. E}}
  \textbf{\bibinfo{volume}{68}},
  \bibinfo{pages}{020401}
 (\bibinfo{year}{2003}).


\bibitem[{\citenamefont{}()}]{Ref11}
  \bibinfo{author}{\bibfnamefont{M.}~\bibnamefont{Ben Amar}},
  \bibinfo{Jurnal}{{Phys. Rev. A.}}
  \textbf{\bibinfo{volume}{44}},
  \bibinfo{pages}{3673}
 (\bibinfo{year}{1991}).


\bibitem[{\citenamefont{}()}]{Ref12}
  \bibinfo{author}{\bibfnamefont{M.}~\bibnamefont{Ben Amar}},
  \bibinfo{Jurnal}{{Phys. Rev. A.}}
  \textbf{\bibinfo{volume}{43}},
  \bibinfo{pages}{5724}
 (\bibinfo{year}{1991}).


\bibitem[{\citenamefont{}()}]{Ref13}
  \bibinfo{author}{\bibfnamefont{A.}~\bibnamefont{Arneodo}},
  \bibinfo{author}{\bibfnamefont{J.}~\bibnamefont{Elezgaray}},
  \bibinfo{author}{\bibfnamefont{M.}~\bibnamefont{Tabard}},
  \bibinfo{author}{\bibfnamefont{F.}~\bibnamefont{Tallet}},
  \bibinfo{Jurnal}{{Phys. Rev. E.}}
  \textbf{\bibinfo{volume}{53}},
  \bibinfo{pages}{6200}
 (\bibinfo{year}{1996}).


\bibitem[{\citenamefont{}()}]{Ref14}
  \bibinfo{author}{\bibfnamefont{L.M.}~\bibnamefont{Sander}},
  \bibinfo{author}{\bibfnamefont{E.}~\bibnamefont{Somfai}},
  \bibinfo{Jurnal}{{Chaos}}
  \textbf{\bibinfo{volume}{15(2)}},
  \bibinfo{pages}{26109}
 (\bibinfo{year}{2005}).


\bibitem[{\citenamefont{}()}]{Ref15}
  \bibinfo{author}{\bibfnamefont{D.A.}~\bibnamefont{Kessler}},
  \bibinfo{author}{\bibfnamefont{Z.}~\bibnamefont{Olami}},
  \bibinfo{author}{\bibfnamefont{J.}~\bibnamefont{Oz}},
  \bibinfo{author}{\bibfnamefont{I.}~\bibnamefont{Procaccia}},
  \bibinfo{author}{\bibfnamefont{E.}~\bibnamefont{Somfai}} and
  \bibinfo{author}{\bibfnamefont{L.M.}~\bibnamefont{Sander}}
  \bibinfo{Jurnal}{{Phys. Rev. E.}}
  \textbf{\bibinfo{volume}{57}},
  \bibinfo{pages}{6913}
 (\bibinfo{year}{1998}).


\bibitem[{\citenamefont{}()}]{Ref16}
  \bibinfo{author}{\bibfnamefont{F.}~\bibnamefont{Mohammadi}},
  \bibinfo{author}{\bibfnamefont{A.A.}~\bibnamefont{Saberi}},
  \bibinfo{author}{\bibfnamefont{S.}~\bibnamefont{Rouhani}},
  \bibinfo{Jurnal}{{J. Phys.: Condens. Matter}}
  \textbf{\bibinfo{volume}{21}},
  \bibinfo{pages}{375110}
 (\bibinfo{year}{2009}).

\end{thebibliography}
\end{document}